\documentclass[aps,prl,preprint,showpacs]{revtex4}
\usepackage{amsfonts}
\usepackage{amsmath}
\usepackage{mathrsfs}
\begin{document}

\title{Spin and orbital angular momentum of the tensor gauge field}

\author{Xiang-Song Chen$^{1,2,3}$}
\email{cxs@hust.edu.cn}

\author{Ben-Chao Zhu$^1$}

\author{Niall \'O Murchadha$^{4}$}
\email{niall@ucc.ie}

\affiliation{$^1$Department of Physics, Huazhong
University of Science and Technology, Wuhan 430074, China\\
$^2$Joint Center for Particle, Nuclear Physics and
Cosmology,
Nanjing 210093, China\\
$^3$Kavli Institute for Theoretical Physics China, Chinese
Academy of Science, Beijing 100190, China\\
$^4$Physics Department, University College Cork, Cork,
Ireland}

\date{\today}

\begin{abstract}
Following the recent studies of the trickiness in spin and
orbital angular momentum of the vector gauge fields, we
perform here a parallel analysis for the tensor gauge
field, which has certain relation to gravitation. Similarly
to the vector case, we find a nice feature that after
removing all gauge degrees of freedom the angular momentum
of the tensor gauge field vanishes for a stationary system.
This angular momentum also shows a one-parameter invariance
over the infinitely many ways of complete gauge fixing for
the tensor field. The tensor gauge coupling, however, does
exhibit a critical difference from the vector gauge
coupling that it may induce intrinsic interaction terms
into the spatial translation and rotation generators,
leaving none of the ten Poincar\'e generators
interaction-free.

%11.15.-q Gauge field theories
%04.20.Cv Fundamental problems and general formalism
\end{abstract}

\maketitle

Recently, the old problem of spin and orbital angular
momentum of the gauge field \cite{textbook} has revived
considerably along two lines. One is the usage of photon
orbital angular momentum in laser beams \cite{Enk07}, the
other is the study of gluon contribution to the nucleon
spin \cite{SpinReview}. The trickiness in these studies is
that the gauge degrees of freedom make it hard to
unambiguously construct a canonical quantity like spin, and
much controversy arose
\cite{Chen08,Chen09,Chen11A,Waka10,Cho10,Lead11,Hatt11}. In
the debate of how to properly define a meaningful spin and
angular momentum for the gauge field, Chen {\it et al}
found a nice feature that the angular momentum of the
vector gauge field can be made vanishing for a stationary
system \cite{Chen11A}. This feature is physically
reasonable and leads to simple pictures of spin structure
for atoms and heavy hadrons \cite{Chen11A}, thus can serve
as a guidance or criteria in proper identification of the
angular momentum for the gauge field. In this paper, we
perform a parallel analysis for the tensor gauge field, and
discuss the remarkable similarities and differences in
comparison to the vector case.

We consider a symmetric tensor field $h_{\mu\nu}$ with a
linear gauge transformation
\begin{equation}
h_{\mu\nu}(x)\to h'_{\mu\nu}(x)=h_{\mu\nu}(x)+\partial_\mu
\xi_\nu(x) +\partial_\nu \xi _\mu(x),\label{gaugetrans}
\end{equation}
where $\xi_\mu(x)$ are four arbitrary gauge parameters. We
consider a general model with $h_{\mu\nu}$ coupled to an
external conserved source $T^{\mu\nu}(x)$, and require the
model be invariant under the gauge transformation in
(\ref{gaugetrans}). Restricted to quadratic terms in first
derivatives, the Lagrangian density of such a model is
essentially unique up to irrelevant total divergences
\cite{Padm08}:
\begin{eqnarray}
\mathscr{L}=\frac 14 (\partial_\mu h^\alpha_{~\alpha}
\partial^\mu h^\beta_{~\beta} -\partial_\mu
h_{\alpha\beta} \partial^\mu h^{\alpha\beta} +2\partial_\mu
h^{\mu\alpha}\partial^\nu h_{\nu\alpha}\nonumber \\
 -2 \partial_\mu h^\alpha_{~\alpha}
\partial_\nu h^{\mu\nu}) +\frac \kappa 2 h_{\mu\nu} T^{\mu\nu}.
\label{Lagr}
\end{eqnarray}
This can be regarded as the weak-field limit of Einstein's
general relativity, with $h_{\mu\nu}$ the metric
perturbation and $T^{\mu\nu}$ the energy-momentum tensor of
matter. But in this paper we just consider a most general
case, and do not assign any specific physical contents to
$h_{\mu\nu}$ and $T^{\mu\nu}$.

Given the Lagrangian, we can proceed to construct the
angular momentum of the tensor gauge field $h_{\mu\nu}(x)$.
We take the canonical expression
\begin{eqnarray}
J_{ij}=\int d^3x \frac {\partial \mathscr{L}}{\partial \dot
h^{\mu\nu}}\left[( x_j\partial_i -x_i\partial_j)
h^{\mu\nu}-i(\Sigma_{ij})^{\mu\nu}_{~~\alpha\beta}h^{\alpha\beta}\right].
\end{eqnarray}
[Conventions: an over dot denotes time derivative, Greek
indices run 0-3, Latin indices run 1-3. Summation is
assumed for repeated indices, even when two spatial indices
are both upstairs or downstairs. This would cause no
trouble since we take the metric $\eta_{\mu\nu}={\rm
diag.}(-1,1,1,1)$.] $\frac {\partial \mathscr{L}}{\partial
\dot h^{\mu\nu}}\equiv \Pi_{\mu\nu}$ is the momentum
conjugate of $h^{\mu\nu}$, and
$(\Sigma_{ij})^{\mu\nu}_{~~\alpha\beta}$ is the spin matrix
governing the Lorentz transformation of $h^{\mu\nu}$. The
angular momentum pseudovector is $J_k=\frac 12
\epsilon_{ijk} J_{ij}$, which acts as the rotation
generator.

Under an infinitesimal Lorentz transformation,
\begin{equation}
x'^\mu =\Lambda^\mu_{~\nu}x^\nu,~
\Lambda^\mu_{~\nu}=\delta^\mu_{~\nu}+\omega^\mu_{~\nu},~\omega^{\mu\nu}=-\omega^{\nu\mu},
\end{equation}
the tensor field transforms as
\begin{equation}
h'^{\mu\nu}=\Lambda^\mu_{~\alpha}\Lambda^\nu_{~\beta}h^{\alpha\beta}\simeq
h^{\mu\nu}+(\delta^\mu_{~\alpha}\omega^\nu_{~\beta}+
\delta^\nu_{~\beta}\omega^\mu_{~\alpha})h^{\alpha\beta}.
\end{equation}
Casting the field variation into the form
\begin{equation}
\delta h^{\mu\nu} =\frac i2
\omega^{\rho\sigma}(\Sigma_{\rho\sigma})^{\mu\nu}_{~~\alpha\beta}h^{\alpha\beta},
\end{equation}
we can read out the spin matrix to be
\begin{equation}
i(\Sigma_{\rho\sigma})^{\mu\nu}_{~~\alpha\beta}=
\delta^\mu_{~\alpha}(\delta^\nu_{~\rho}\eta_{\sigma\beta}
-\delta^\nu_{~\sigma}\eta_{\rho\beta})
+\delta^\nu_{~\beta}(\delta^\mu_{~\rho}\eta_{\alpha\sigma}-
\delta^\mu_{~\sigma}\eta_{\alpha\rho}).
\end{equation}

The momentum conjugates are
\begin{subequations}
\begin{eqnarray}
\Pi_{00}&=&\frac 12 \partial_ih_{0i},\\
\Pi_{0i}&=&\frac 12 (\partial_i h_{00}-\partial_i h +2
\partial_j h_{ij}),\\
\Pi_{ij}&=&\frac 12 [\dot h_{ij} +\delta_{ij}(\partial_k
h_{0k}- \dot h)].
\end{eqnarray}
\end{subequations}
Here $h\equiv h_{ii}$ is the spatial trace. [We remark that
we have identified $\dot h_{0i}$ with $\dot h_{i0}$, but
not $\dot h_{ij}$ with $\dot h_{ji}$, thus we are going to
sum over both $(ij)$ and $(ji)$, but not $(i0)$.]

It is now straightforward to compute the angular momentum
tensor $J_{ij}$. The spin part is found to be
\begin{eqnarray}
S_{ij}&\equiv& \int d^3x \Pi_{\mu\nu}
i(\Sigma_{ij})^{\mu\nu}_{~~\alpha\beta}h^{\alpha\beta}\nonumber \\
&=&\frac 12 \int d^3x[2(\dot h_{jk}h_{ik} -\dot h_{ik}h_{jk}) \nonumber \\
&+& h_{0j}(2\partial_k h_{ik}+\partial_i h_{00} -\partial_i
h)\nonumber \\
&-& h_{0i}(2\partial_k h_{jk}+\partial_j h_{00} -\partial_j
h)]. \label{Sh}
\end{eqnarray}

The orbital part is found to be
\begin{eqnarray}
L_{ij}&\equiv&\int d^3x \Pi_{\mu\nu}(x_j\partial_i
- x_i\partial_j)h^{\mu\nu}\nonumber \\
&=&\frac 12 \int d^3 x  [\dot h_{kl} (x_j\partial_i
- x_i\partial_j) h_{kl} \nonumber \\
&-&(2\partial_l h_{kl} +\partial_k h_{00}-\partial_k
h)(x_j\partial_i - x_i\partial_j)h_{0k}\nonumber \\
&+&\partial_k h_{0k}(x_j\partial_i
- x_i\partial_j)h_{00}\nonumber \\
&+&(\partial_k h_{0k} -\dot h) (x_j\partial_i -
x_i\partial_j) h ]. \label{Lh}
\end{eqnarray}

For comparison, we quote the corresponding expressions for
the vector gauge field $A^\mu$, denoted by a superscript
$^A$.
\begin{eqnarray}
{\mathscr L}^A&=& \frac 12 (\partial_\mu A_\nu
\partial^\nu A^\mu-\partial_\mu A_\nu\partial^\mu
A^\nu)+eA_\mu j^\mu ,\\
S^A_{ij}&=&\int d^3x [\dot A_j A_i-\dot A_i A_j\nonumber  \\
&+&\partial_j A^0 A_i-\partial_i A^0 A_j] ,\\
L^A_{ij}&=&\int d^3x [\dot A_k(x_j\partial_i-x_i\partial_j)
A_k\nonumber \\
&+&\partial_kA^0(x_j\partial_i-x_i\partial_j) A_k] .
\end{eqnarray}

We note the following similarities and differences between
$S_{ij}$, $L_{ij}$, $J_{ij}$ and $S^A_{ij}$, $L^A_{ij}$,
$J^A_{ij}\equiv S^A_{ij}+ L^A_{ij}$:

(i) They are all gauge-dependent. Such a gauge-dependence
has long obscured the physical meanings of photon spin and
orbital angular momentum \cite{textbook}.

(ii) They all contain terms that involve no time
derivative, and thus can survive for a stationary
configuration. We will loosely call these terms ``static'',
though they can certainly be time-dependent as well.

(iii) $S_{ij}$ and $L_{ij}$ appear much more complicated
than $S^A_{ij}$ and $L^A_{ij}$. A major cause is that $\dot
A^0$ drops out in ${\mathscr L}^A$, but $\dot h_{00}$ and
$\dot h_{0i}$ survive in ${\mathscr L}$ (though not
quadratically).

(iv) $L_{ij}$ contains a novel trace term with $\dot h$.

In common textbooks on classical electrodynamics, it is
popular to discuss angular momentum of a static
electromagnetic field. But this notion is really peculiar,
since the electromagnetic field is massless, and should
possess no momentum when ``not moving''. Indeed, it was
show in Ref. \cite{Chen11A} that the total angular momentum
of the vector gauge field can be constructed to vanish
identically for a stationary system. When adopting the
above gauge-dependent expressions, this feature occurs {\em
in and only in} the Coulomb gauge. This phenomenon is
fairly delicate and needs some elaboration. First, the
static terms in $S^A_{ij}$ and $L^A_{ij}$ sum to be
\begin{eqnarray}
\int d^3x [\partial_j A^0 A_i-\partial_i A^0
A_j+\partial_kA^0(x_j\partial_i-x_i\partial_j)
A_k]\nonumber \\
=-\int d^3x A^0 (x_j\partial_i-x_i\partial_j) (\partial_k
A_k).
\end{eqnarray}
Thus, in Coulomb gauge, $\vec \partial \cdot \vec A=0$, the
static terms in $J^A_{ij}$ vanish and $J^A_{ij}$ simplifies
to
\begin{equation}
^C J^A_{ij}=\int d^3x [\dot A_j A_i-\dot A_i A_j+\dot
A_k(x_j\partial_i-x_i\partial_j) A_k]^C. \label{JA}
\end{equation}
The superscript $^C$ denotes imposition of Coulomb gauge.
Each term now contains a time-derivative. However, there
remains a gap to claim that $^C J^A_{ij}$ vanishes for a
stationary system: The stationary condition only means that
the gauge-invariant physical observables (like the electric
current $j^\mu$ or electromagnetic field
$F^{\mu\nu}=\partial^\mu A^\nu-\partial^\nu A^\mu$) are
time-independent, while the gauge-potential $A^\mu$ may
contain spurious (nonphysical) time-dependence \cite{note}.
This gap is closed by noting that in Coulomb gauge $A^\mu$
can be expressed in terms of $F^{\mu\nu}$ \cite{Chen11d}:
\begin{equation}
^C A^\mu =\frac 1 {\vec
\partial^2}
\partial_i F^{i\mu}. \label{Ac}
\end{equation}
Hence, in Coulomb gauge, $A^\mu$ is time-independent if
$F^{\mu\nu}$ is, and Eq. (\ref{JA}) dictates that $^C
J^A_{ij}$ vanishes for a stationary system.

Eq. (\ref{Ac}) shows a delicate {\em dual} relation between
gauge-fixed and gauge-invariant expressions, as we
carefully discussed in \cite{Chen11T}: If one solely looks
at the right-hand-side of Eq. (\ref{Ac}), one can in
principle forget all about Coulomb gauge, and define $\frac
1 {\vec \partial^2} \partial_i F^{i\mu}$ as a
gauge-invariant ``physical field'' $\hat A^\mu$. (Certainly
this $\hat A^\mu$ agrees with $A^\mu$ in Coulomb gauge, and
its spatial part $\hat A_i$ is just the transverse field
$A^\perp _i=A_i -\partial_i \frac 1{\vec\partial^2} \vec
\partial \cdot \vec A$.) Analogously, in Eq. (\ref{JA}), if
one substitutes $^C A_i$ with the explicit expression in
Eq. (\ref{Ac}) (this is equivalent to replacing $^C A_i$
with $\hat A_i$), then in the final expression for
$J^A_{ij}$ one can again forget all about Coulomb gauge,
and regard the expression as the definition of a
gauge-invariant $\hat J^A_{ij}$, which then vanishes
identically for a stationary system, in any gauge for
$A^\mu$.

As we explained in \cite{Chen11T}, such a dual relation is
only possible if the gauge-fixing is indeed complete. The
special role of Coulomb gauge (for a vector field) is
exactly that it completely removes the gauge degrees of
freedom under a trivial boundary condition. Thus the
finding of Ref. \cite{Chen11A} is that the physical degrees
of freedom of the vector gauge field contribute no angular
momentum for a stationary system. Remarkably, we find that
the same feature can be demonstrated for the much more
complicated tensor gauge field.

Following the hints from the vector case, we look at the
canonical expressions in Eqs. (\ref{Sh}) and (\ref{Lh}),
and examine their properties by applying a complete gauge
constraint on $h_{\mu\nu}$. Such a complete tensor gauge
condition, however, is not unique \cite{Chen11T}. It can
take a general form:
\begin{equation}
\partial_i h_{0i} +a\partial_0 h_{ii} =0, ~~~
\partial_i h_{ji} +b\partial_j h_{ii} =0. \label{ab}
\end{equation}
The parameters $a,b$ can take any value except $b=-1$,
which is excluded because $\partial_i h_{ji}-\partial_j
h_{ii}$ has a gauge-invariant divergence and thus is unable
to fix any gauge. That the constraints in (\ref{ab}) make a
complete gauge condition can be seen in two ways
\cite{Chen11T}. First, (\ref{ab}) permits no more gauge
freedom; and second, the gauge-transformation parameter
$\xi_\mu$ that brings $h_{\mu\nu}$ to the gauge (\ref{ab})
is unique. The special properties of some particular
choices of $a,b$ are discussed in \cite{Chen11T}. Till the
end of our derivation, we will see an interesting
one-parameter gauge-invariance for the angular momentum of
the tensor gauge field.

Taking the gauge condition in (\ref{ab}), and applying some
slight algebra, we find the simplified expressions:
\begin{eqnarray}
S_{ij}^{(ab)}&=& \frac 12\int d^3x[2(\dot h_{jk}h_{ik}
-\dot h_{ik}h_{jk}) \nonumber \\
&+&(h_{00}-(2b+1)h)(\partial_j h_{0i}-\partial_i
h_{0j})]^{(ab)}.\\
L_{ij}^{(ab)} &=&\frac 12\int d^3 x  [\dot h_{kl}
(x_j\partial_i
- x_i\partial_j) h_{kl} \nonumber \\
&-&(2ab+2a+1) \dot h (x_j\partial_i
- x_i\partial_j) h \nonumber \\
&-&(h_{00}-(2b+1)h)(\partial_j h_{0i}-\partial_i
h_{0j})]^{(ab)}.
\end{eqnarray}
The superscript $^{(ab)}$ denotes imposition of the gauge
in (\ref{ab}). The static terms cancel exactly between
$S_{ij}^{(ab)}$ and $L_{ij}^{(ab)}$, and the total
$J_{ij}^{(ab)}$ becomes
\begin{eqnarray}
J_{ij}^{(ab)}&=&\frac 12 \int d^3x[2(\dot h_{jk}h_{ik}
-\dot h_{ik}h_{jk}) \nonumber \\
&+&\dot h_{kl}(x_j\partial_i - x_i\partial_j) h_{kl}\nonumber \\
&-&(2ab+2a+1) \dot h (x_j\partial_i - x_i\partial_j) h
]^{(ab)}. \label{Jh-ab}
\end{eqnarray}
Similarly to Eq. (\ref{JA}), each term in $J_{ij}^{(ab)}$
contains a time-derivative. But as we remarked above, to
conclude that $J_{ij}^{(ab)}$ vanishes for a stationary
system, we still need to show that $h_{ij}^{(ab)}$ cannot
induce spurious time-dependence. This property can be
inferred from our recent careful examination of tensor
gauge conditions \cite{Chen11d,Chen11T}:
\begin{subequations}
\label{ab-CZ}
\begin{eqnarray}
h_{ij}^{(ab)}&=&f_{ij} -\frac {1+2b}{2(1+b)}\frac 1 {\vec
\partial ^2} (\partial_i\partial_j f_{kk}), \\
h_{0j}^{(ab)}&=&f_{0j} -\frac {1+a+b}{2(1+b)}\frac 1 {\vec
\partial ^2} (\partial_0\partial_j f_{kk}),\\
h_{00}^{(ab)}&=&f_{00} -\frac {1+2a}{2(1+b)}\frac 1 {\vec
\partial ^2} (\partial_0^2 f_{kk}).
\end{eqnarray}
\end{subequations}
Here $f_{\mu\nu}\equiv 2 \frac 1 {\vec\partial^2} R_{\mu
ii\nu}$, and $R_{\mu\rho\sigma\nu}$ is the linearized
Riemann curvature. For completeness and future reference,
we have displayed all ten components of
$h_{\mu\nu}^{(ab)}$.

Eqs. (\ref{ab-CZ}) indicate clearly that
$h_{\mu\nu}^{(ab)}$ is time-independent if the
gauge-invariant $R_{\rho\sigma\mu\nu}$ is, hence
$J_{ij}^{(ab)}$ vanishes for a stationary system. We thus
proved the same nice feature as in the vector case that the
physical degrees of freedom of the tensor gauge field carry
no static angular momentum. Moreover, as in the vector
case, one can also define the right-hand-side of Eqs.
(\ref{ab-CZ}) as a gauge-invariant physical field $\hat
h_{\mu\nu}$, and forget all about the gauge in (\ref{ab}).
With this $\hat h_{\mu\nu}$, one can define a
gauge-invariant $\hat J_{ij}$ by replacing $h_{ij}$ in Eq.
(\ref{Jh-ab}) with $\hat h_{ij}$, and this $\hat J_{ij}$
vanishes identically under the stationary condition, in any
gauge for $h_{\mu\nu}$.

The expression in Eq. (\ref{Jh-ab}) is not yet the final
story, as it has not reached the art of Eq. (\ref{JA}),
where $^C A_i=A_i^\perp$ represents the two dynamical
(propagating) components of the vector field. In Eq.
(\ref{Jh-ab}) the trace $h^{(ab)}$ is non-dynamical, as
revealed by its equation of motion \cite{Chen11T}:
\begin{equation}
\vec\partial ^2 h^{(ab)}=-\frac \kappa {1+b} T_{00}.
\label{h}
\end{equation}
The instantaneous feature of the Laplacian operator $\vec
\partial ^2$ means that $h^{(ab)}$ is completely dictated by the
source. Namely, $h^{(ab)}$ is not an independent dynamical
quantity that can propagate. An important implication of
this fact is that $h_{ij}^{(ab)}$, with a nonzero trace, is
not fully dynamical either. Furthermore, by the gauge
condition in (\ref{ab}), the spatial divergence of
$h_{ij}^{(ab)}$ is non-dynamical as well. To get the purely
dynamical component of $h_{ij}^{(ab)}$, we thus need to
extract its transverse-traceless (TT) part $h^{TT}_{ij}$
\cite{ADM}. This $h^{TT}_{ij}$ is completely invariant
under gauge transformation in (\ref{gaugetrans}). It is the
counter part of $A_i^\perp$ for the vector field. Ref.
\cite{Chen11T} gives how $h^{TT}_{ij}$ relates to
$h_{ij}^{(ab)}$:
\begin{equation}
h_{ij}^{(ab)}=h^{TT}_{ij}+\frac {1+b}2 \delta _{ij}
h^{(ab)}-\frac {1+3b} 2 \frac 1{\vec\partial^2}
\partial_i\partial_j h^{(ab)}.
\end{equation}
Inserting this into Eq. (\ref{Jh-ab}), and using Eq.
(\ref{h}) for $h^{(ab)}$, we obtain:
\begin{eqnarray}
J_{ij}^{(ab)}&=&\frac 12 \int d^3x[2(\dot
h^{TT}_{jk}h^{TT}_{ik}
-\dot h^{TT}_{ik}h^{TT}_{jk}) \nonumber \\
&+&\dot h^{TT}_{kl} (x_j\partial_i-x_i\partial_j)
h^{TT}_{kl}
\nonumber \\
&-&\kappa^2 (\frac 12 + 2\frac {a-b}{1+b}) \frac {\dot
T_{00}}{\vec\partial ^2}
(x_j\partial_i-x_i\partial_j)\frac{T_{00}} {\vec\partial
^2}]. \label{J-final}
\end{eqnarray}

The last term in Eq. (\ref{J-final}) is a bit special and
calls for attention. It contains all the dependence of
$J_{ij}^{(ab)}$ on the two gauge parameters $a,b$, but
through a single factor $(\frac 12 + 2\frac {a-b}{1+b})$.
$J_{ij}^{(ab)}$ thus possesses a one-parameter invariance:
$(\frac 12 + 2\frac {a-b}{1+b})$ can take a universal value
$\lambda$ for any $a=b+(\frac \lambda 2-\frac 14) (1+b)$.
One interesting example is $a=b$, which gives
$\lambda=\frac 12$. The most attractive choice might be
$\lambda=0$ for any $a=\frac 14 (3b-1)$. With $\lambda=0$,
Eq. (\ref{J-final}) reduces to the same form as for a free
field in the absence of source, and mimics exactly Eq.
(\ref{JA}), whose form is unaltered by the presence of
source.

The gauge with $a=\frac 14 (3b-1)$, however, is not
necessarily consistent with quantum Lorentz invariance. As
Weinberg elaborated in \cite{Wein65}, by canonical
quantization of tensor gauge field with only physical
degrees of freedom, Lorentz invariance of S-matrix requires
a delicate matching between the Hamiltonian and propagator.
This matching can be achieved in some particular gauge.
E.g., Weinberg found $a=-\frac 23$ and $b=-\frac 13$, which
however does not fall into the class of $a=\frac 14
(3b-1)$.

With $(\frac 12 + 2\frac {a-b}{1+b}) \neq 0$, the last term
in Eq. (\ref{J-final}) is then intrinsic and novel. The
appearance of the coupling constant $\kappa$ means that
this term represents an interaction effect. On the other
hand, it is entirely expressed in terms of the source and
should apparently be counted as part of the source angular
momentum. One should note, however, that such a term is
absent for a free source which does not couple to the
tensor gauge field. Therefore, the presence of such a term
seems to indicate that, unlike the vector gauge coupling in
the standard model of particle physics, the tensor gauge
coupling induces extra term into the angular momentum of
the system. In other words, the tensor gauge coupling
modifies the rotation generator of the system.

Exactly analogous situation can be demonstrated for the
spatial translation generator (or the momentum) of the
tensor gauge field:
\begin{eqnarray}
\vec P&\equiv& -\int d^3x \Pi_{\mu\nu} \vec
\partial h^{\mu\nu} =-\frac 12 \int d^3x [\dot h_{kl}
\vec \partial h_{kl} -\dot h \vec \partial h \nonumber
\\&&+2\partial_k h_{0k} \vec \partial h -2\partial_l
h_{kl}\vec\partial h_{0k}].
\end{eqnarray}
Here we also find static terms, which seem to imply that a
stationary tensor field can possess momentum. But after
applying the constraint (\ref{ab}) to remove all gauge
degrees of freedom, $\vec P$ simplifies to
\begin{eqnarray}
\vec P^{(ab)}=-\frac 12 \int d^3x [\dot h_{kl} \vec
\partial h_{kl}
-(2ab+2a+1) \dot h \vec \partial h]^{(ab)} \nonumber\\
= -\frac 12 \int d^3x [\dot h^{TT}_{kl} \vec \partial
h^{TT}_{kl} -\kappa^2 (\frac 12 + 2\frac {a-b}{1+b}) \frac
{\dot T_{00}}{\vec\partial ^2}\vec \partial \frac{T_{00}}
{\vec\partial ^2}]. \label{P-final}
\end{eqnarray}
This is clearly zero for a stationary system, showing that
a static, massless tensor gauge field possesses no physical
momentum. The second expression in Eq. (\ref{P-final})
results from extracting the $TT$ part and using Eq.
(\ref{h}), and we find the same factor $(\frac 12 + 2\frac
{a-b}{1+b})$ as in Eq. (\ref{J-final}).

In comparison, the momentum expression for the vector gauge
field is
\begin{equation}
\vec P^A\equiv -\int d^3x \frac{\partial {\mathscr
L}^A}{\partial \dot A_k} \vec \partial A_k=-\int d^3x (\dot
A_k+\partial_k A^0)\vec \partial A_k.
\end{equation}
In Coulomb gauge, this reduces to
\begin{eqnarray}
^C \vec P^A=-\int d^3x [\dot A_k \vec \partial A_k]^C
=-\int d^3x \dot A^\perp_k \vec \partial A^\perp_k.
\end{eqnarray}

We see again that unlike the vector gauge coupling, the
tensor gauge coupling induces an interaction term in the
spatial translation generator. This seems to imply that
with the tensor gauge coupling the ten Poincar\'e
generators are all ``bad'' (in the sense of containing
interaction), while for the vector gauge coupling only four
generators (for time translation and Lorentz boost) are
bad, and six generators (for spatial translation and
rotation) remain ``good'' (interaction-free) \cite{Wein95}.
This implication, however, is not decisive, since we have
not included dynamical part for the source; and the subject
needs further careful investigation.

This work is supported by the China NSF Grants 10875082 and
11035003. X.S.C. is also supported by the NCET Program of
the China Education Department.


\begin{thebibliography}{99}
\bibitem{note} For example, the Lorentz gauge $\vec
\partial \cdot \vec A +\partial_t A^0=0$ does not strictly
reduce to Coulomb gauge for a stationary system, because
one can assign a time-dependent $A^0$ in Lorentz gauge
while leaving $F^{\mu\nu}$ time-independent.
\end{thebibliography}
\end{document}